\documentstyle[twocolumn,prl,aps]{revtex}
\input epsf

\draft

\begin{document}

\title{Existence of an omni-directional photonic band gap in one-dimensional periodic dielectric structures~\thanks{Submitted to Physical Review Letters on June 1, 1998; resubmitted on October 12, 1998.}}

\author{Dmitry N. Chigrin~\cite{DChigrin}}
\address{Universit{\" a}t Gesamthochschule Essen, Fachbereich Physik,
45117 Essen, Germany}

\author{Andrei V. Lavrinenko}
\address{Belarusian State University, Department of  Physics,
Fr. Skarina Ave. 4, 220080 Minsk, Belarus}

\author{\parbox[t]{5.5in}{\small
It is shown that total reflection for all incident angles does not require a two- or three-dimensional photonic crystal. We demonstrate that a one-dimensional photonic crystal can exhibit total omni-directional reflection for any incident wave within some frequency region. The formation of the omni-directional gap is discussed and a wide range of realistic fabrication parameters is proposed.
}}
\maketitle
\normalsize

Since the last decade the investigations of photonic band gap materials~\cite{Basic}, so-called photonic crystals, have been attracting close attention due to the various optoelectronics' applications~\cite{PBGs,Joannopoulos} that will revolutionize photonics. It is very important to design and fabricate artificial microstructures possessing an absolute and complete three-dimensional (3D) photonic band gap, i.e., structures forbidding electromagnetic wave propagation for all polarizations and directions. One of the promising properties of such a structure is that a thick stack of finite thickness behaves as a photonic insulator exhibiting total omni-directional reflection within some frequency region. For the moment most of the efforts are related to synthesis of 2D and 3D periodic structures~\cite{Bogomolov} and the applications are restricted by technological difficulties.

In this Letter, we report that total omni-directional reflection can be obtained with a one-dimensional (1D) photonic crystal~\cite{Patent}. The high reflectivity of 1D periodic structures (so-called reflective dielectric coatings) is a well-known phenomenon that has been studied for a long time~\cite{Joannopoulos,Born,Yeh}. However, to our knowledge, the required conditions to obtain total reflection for a wide range of incident angles have not yet been reported. The 1D periodic structures are much easier to fabricate than 2D and 3D ones. Therefore, the existence of  total omni-directional reflection from 1D systems permits to design extremely cheap and effective photonic insulator, tuned to the needed frequency by changing the period of the structure.

\begin{figure}[tb]
\noindent \epsfbox{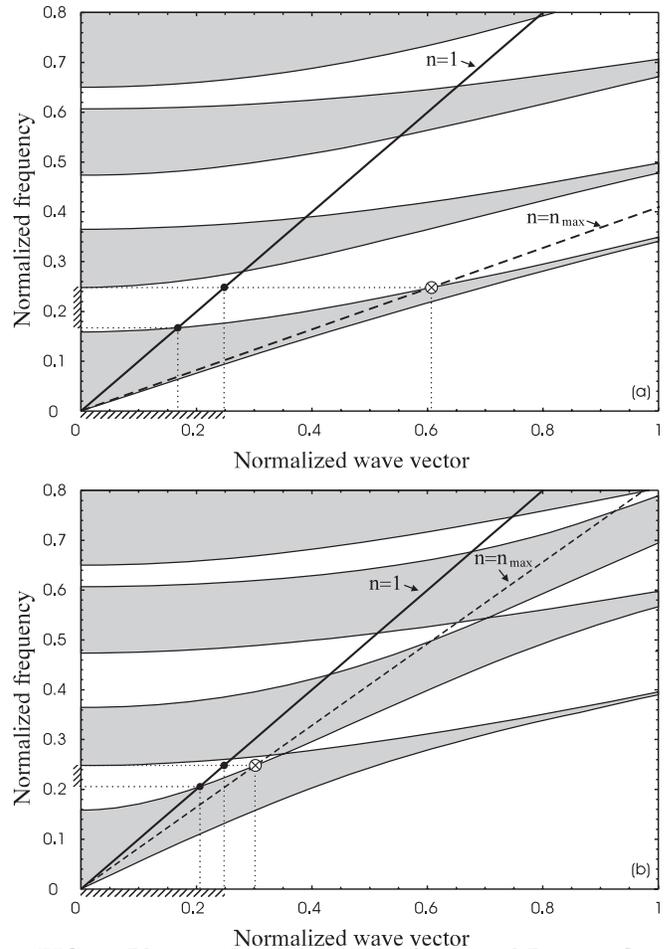}
\caption{Photonic band structure of a typical Bragg reflector for TE~(a) and TM~(b) polarizations. The frequency and the tangential component of the wave vector are defined to be normalized as $\omega \Lambda /2\pi c$  and $\left|{\bf k}_\perp\right|\Lambda/2\pi$, respectively. Here, $n_1=1.4$, $n_2=3.4$ and $\eta=1.0$.}
\label{fig:fig1}
\end{figure}

For the present study, we focus on the simplest type of 1D photonic crystal, which is a Bragg reflector, a periodic array of alternating layers with low, $n_1$, and high, $n_2$, indices of refraction and  the thicknesses $d_1$ and $d_2$, respectively. The period of the structure is $\Lambda=d_1+d_2$. The plane electromagnetic wave is considered to illuminate the boundary of a semi-infinite photonic crystal under the angle, $\alpha$, from semi-infinite homogeneous medium with refraction index, $n$. The photonic band structure for a typical Bragg reflector with $n_1=1.4$, $n_2=3.4$ and a filling factor $\eta=d_2/d_1=1.0$ is shown in the figures~\ref{fig:fig1}(a) and~\ref{fig:fig1}(b) for TE and TM waves, respectively. The band structure has been calculated using the analytical form of the dispersion equation derived for an infinite medium (see e.g. \cite{Yeh}). It is important to note that infinite and semi-infinite photonic crystals have the same band structure~\cite{Zolla}, the only difference is the existence of surface modes in the case of semi-infinite structure. The main feature of all 1D photonic crystals is that although forbidden gaps exist for most given values of the tangential component of the wave vector (${\bf k}_\perp$), there is not an absolute nor complete photonic band gap if all possible values of the tangential component of the wave vector are considered~\cite{Joannopoulos,Yeh} (Fig.~\ref{fig:fig1}). 
However, when a plane wave with the wave vector $\left|{\bf k}\right|=n\omega/c$ illuminates a 1D semi-infinite photonic crystal, the tangential component of the wave vector remains constant throughout the crystal and equals to $\left|{\bf k}_\perp\right|=(n\omega/c)\sin{\alpha}$. Here, $\omega$ is the frequency and $c$ is the speed of light in vacuum. The tangential component of the wave vector lies between zero for normal incidence ($\alpha=0^{\circ}$) and $n\omega/c$ for grazing incidence ($\alpha=90^{\circ}$). Therefore, only values of the tangential component of the wave vector, which are inside the range, need to be considered. For this limited range  (the shaded area on the wave vector axis, Fig.~\ref{fig:fig1}), there is the frequency region (the shaded area on the frequency axis, Fig.~\ref{fig:fig1}) which is completely inside the forbidden gaps of the photonic crystal both for TE and TM polarizations. No propagating mode are allowed in the photonic crystal for any propagating mode in the ambient medium within this frequency region. Therefore, the total reflection occurs for any incident angle and omni-directional photonic band gaps (PBGs) are formed for TE and TM polarizations. A band of surface modes lies below the line $\omega=c\left|{\bf k}_\perp\right|/ n \sin\alpha$~\cite{Joannopoulos} and thus surface modes do not affect the external reflectivity.

\begin{figure}[tb]
\noindent \epsfbox{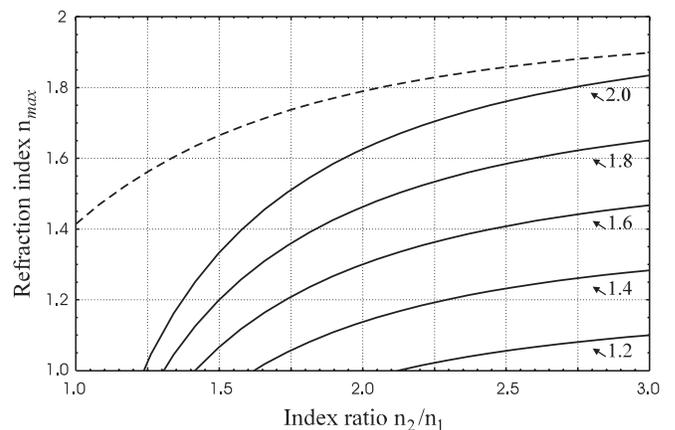}
\caption{Maximum refraction index of the ambient medium, for which an absolute gap is closed up, as a function of the index ratio $\delta n=n_2/n_1$ for different smaller indices $n_1$ and fixed filling factor $\eta=1$. The dashed curve is for  refraction index corresponding to the Brewster angle $\alpha_B$ on the $n_2-n_1$ boundary ($n_{max}=n_2\sin{\alpha_B}$).}
\label{fig:fig2}
\end{figure}

The upper edge of omni-directional PBGs corresponds to the upper edge of the forbidden gap at normal incidence. The lower edge is defined by the intersection of the line $\omega=c\left|{\bf k}_\perp\right|/n$ with the upper edge of the corresponding dielectric band.  In general, the degeneracy between polarizations is lost for off-axis propagation ($\left|{\bf k}_\perp\right| \neq 0$) and the lower edges of the omni-directional PBGs do not coincide for the two fundamental polarizations. However, an absolute omni-directional PBG is formed due to the overlap of the gaps for TE and TM polarizations. For given parameters of the Bragg reflector, the refraction index of an ambient medium may be used to control the width of omni-directional PBGs. Whenas one increases the refraction index, the lower edge of the gaps shifts towards a higher frequency and the gaps width decreases. The maximum width is obtained for  $n=1$ (the solid lines in figure~\ref{fig:fig1}). Omni-directional PBGs are closed up for $n_{max}=c\left|{\bf k}_\perp(\omega_h)\right|/\omega_{h}$, where $\omega_{h}$ is the upper edge of the gaps and $\left|{\bf k}_\perp(\omega_h)\right|$ corresponds to the allowed mode of the photonic crystal for the frequency $\omega_{h}$ (the dashed lines in figure~\ref{fig:fig1}). Due to the Brewster effect on the $n_2-n_1$ boundary, TM forbidden gaps shrink to zero when $\omega=c \left|{\bf k}_\perp\right|/n_2\sin{\alpha_B}$ [Fig.~\ref{fig:fig1}(b)], where $\alpha_B$ is the Brewster angle. Thus, the omni-directional PBG is always narrower and is closed up for a smaller refraction index of the ambient medium for TM polarization than for TE one. To have an absolute omni-directional PBG, an omni-directional PBG must be opened both for TE and TM polarization, that means that the refraction index of the ambient medium must be less than the refraction index corresponding to the Brewster angle $\alpha_B$ ($n<n_2\sin{\alpha_B}$). In figure~\ref{fig:fig2} the maximum refraction index of the ambient medium, for which the absolute gap is closed up,  is presented as a function of the index ratio $\delta n=n_2/n_1$ for various values of a smaller refraction index $n_1$ and a fixed filling factor $\eta=1$. 

\begin{figure}[tb]
\noindent \epsfbox{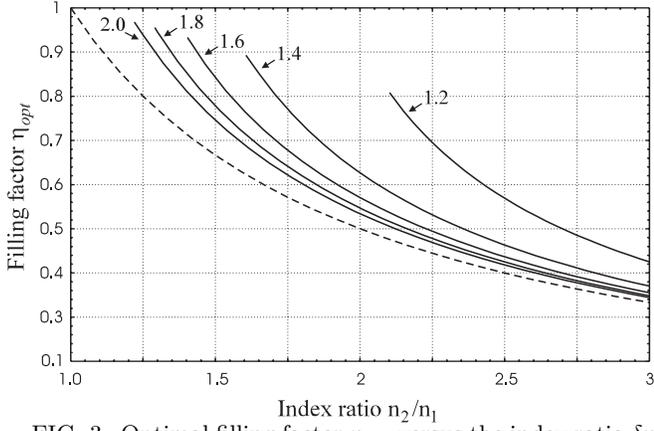}
\caption{Optimal filling factor $\eta_{opt}$ versus the index ratio $\delta n$ for different index ratios $\delta n_0$. The dashed curve corresponds to the filling factor of a quarter-wave stack.}
\label{fig:fig3}
\end{figure}

\begin{figure}[tb]
\noindent \epsfbox{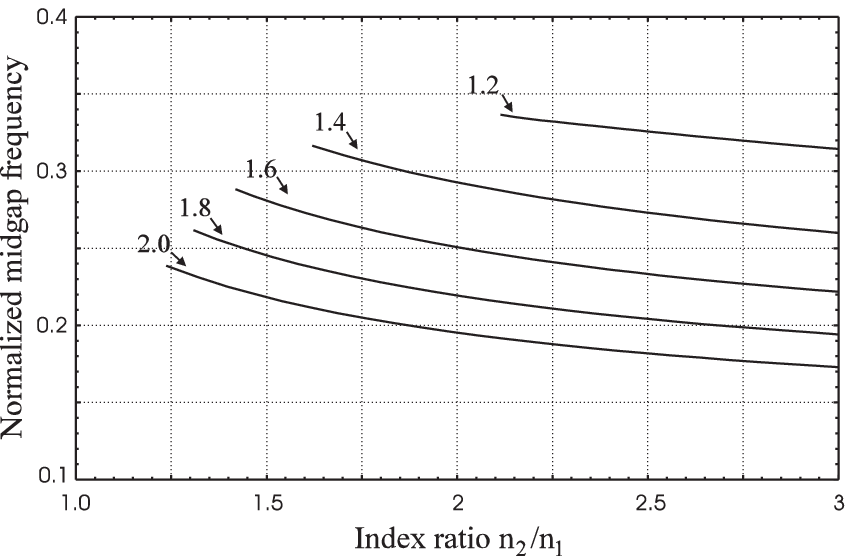}
\caption{Midgap frequency $\omega_0$ versus the  index ratio $\delta n$ for different index ratios $\delta n_0$ for optimal filling factor  $\eta_{opt}$.}
\label{fig:fig4}
\end{figure}

\begin{figure}[tb]
\noindent \epsfbox{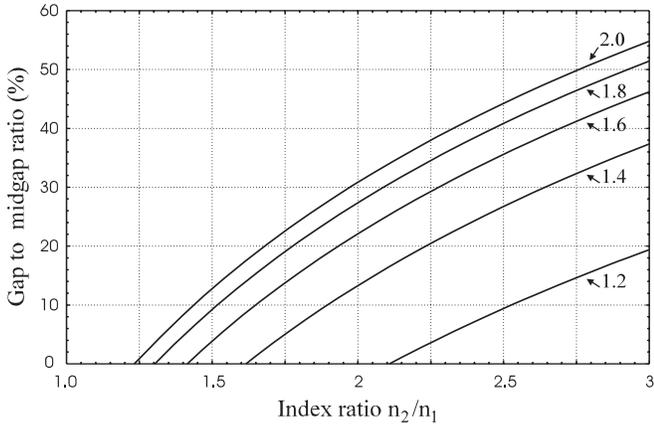}
\caption{Gap to midgap ratio $\Delta \omega/\omega_0$ versus the index ratio $\delta n$ for different index ratios $\delta n_0$ for optimal filling factor  $\eta_{opt}$.}
\label{fig:fig5}
\end{figure}

The filling factor $\eta$ optimizes the gap to midgap ratio $\Delta \omega/\omega_0$ of an absolute omni-directional PBG with respect to the given refraction indices of the layers constituting 1D photonic crystal. Here $\Delta \omega$ is the width of the absolute gap and $\omega_0$ is the midgap frequency. We present the set of contour plots (Fig.~\ref{fig:fig3}-\ref{fig:fig5}) which provides the full information about an absolute omni-directional PBG for given parameters of the Bragg reflector. An optimal filling factor and corresponding midgap frequency are shown in figures~\ref{fig:fig3} and~\ref{fig:fig4}, respectively, as a functions of the index ratio $\delta n$ for different index ratios $\delta n_0=n_1/n$, which is the ratio of smaller refraction index to the refraction index of the ambient medium. The dashed curve in figure~\ref{fig:fig3} corresponds to the filling factor of a quarter-wave stack, which is $\eta_{\lambda /4}=1/\delta n$. Within a given parameter range a quarter-wave stack is not an optimal configuration to reach a maximum gap to midgap ratio for the absolute omni-directional PBG. In figure~\ref{fig:fig5} the optimal gap to midgap ratio is depicted as a function of the index ratio $\delta n$ for different index ratios $\delta n_0$. A wide absolute omni-directional PBG exists for reasonable values of both $\delta n$ and $\delta n_0$. To obtain, for example, an omni-directional gap with a gap to midgap ratio bigger than 5\% the index ratios should be bigger than 1.5 ($\delta n>1.5, \delta n_0>1.5$).  A decrease of one of the index ratios is partially compensated by an increase of the other one.

\begin{figure}[tb]
\noindent \epsfbox{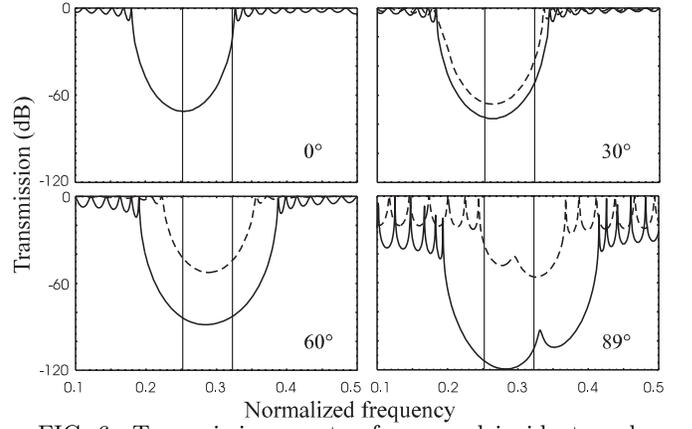}
\caption{Transmission spectra for several incident angles, $\alpha=0^{\circ}$, $30^{\circ}$, $60^{\circ}$, $89^{\circ}$. The solid (dashed) curves are for TE (TM) polarization. The vertical lines on each spectrum indicate an absolute omni-directional photonic band gap. Here, $n=1$, $n_1=1.4$, $n_2=3.4$ and $\eta_{opt}=0.402$. The number of periods is 10.}
\label{fig:fig6}
\end{figure}

\begin{figure}[tb]
\noindent \epsfbox{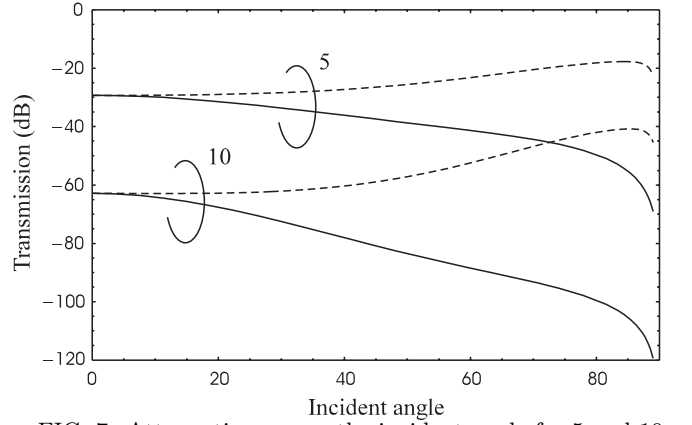}
\caption{Attenuation versus the incident angle for 5 and 10 periods. The solid (dashed) curves are for TE (TM) polarization. Here, $n=1$, $n_1=1.4$, $n_2=3.4$, $\eta_{opt}=0.402$ and $\omega \Lambda /2\pi c=0.287$. }
\label{fig:fig7}
\end{figure}

For example, for a structure with $n=1$, $n_1=1.4$ and $n_2=3.4$ an absolute omni-directional PBG is centered at the normalized frequency $\omega \Lambda /2\pi c=0.287$ with an optimal filling factor $\eta_{opt}=0.402$. The gap to midgap ration is about 24.5\%. The calculated transmission spectra for such a structure with 10 periods are shown in the figure~\ref{fig:fig6}  for several angles of incidence $\alpha$. For calculations of the spectra we used the characteristic matrix method~\cite{Born}. The transmission decreases exponentially inside the gaps with the thickness of the structure. For any given incident angle, the transmission spectra have clear stopbands both for TE and TM polarizations. The midgap frequency shifts to higher frequencies with the incident angle relatively to the midgap frequency corresponding to normal incidence, meanwhile the gaps widths are big enough to possess a sizable intersection of the stopbands at different angle. In the figure~\ref{fig:fig7} the transmission behavior for two structures with 5 and 10 periods is shown against the incident angle for midgap frequency of the absolute omni-directional gap. For TM polarization (dashed curves) there is attenuation minimum near the angle $\alpha=85^{\circ}$ due to the Brewster effect on the first boundary of the stack. For the structure with 10 periods, the minimum over-all obtainable transmission is $-40$~dB. This attenuation is large enough to be used in optical devices. To obtain an absolute omni-directional gap centered at the radiation wavelength $\lambda=0.60 \mu m$, one needs a period of the structure of about $0.172 \mu m$ for the considered parameters. The technological success in thin-film preparation is great enough to fabricate photonic insulators with operating range centered at frequencies from the microwave to the ultraviolet regime.

In conclusion, we have demonstrated the existence of absolute omni-directional photonic band gap in one-dimensional periodic dielectric structures. The evolution of the gap characteristics versus a comprehensive set of the structure parameters have been presented. We have found that for reasonable values of structure parameters ($\delta n>1.5, \delta n_0>1.5$) a relatively large omni-directional gap ($>5\%$) may be obtain, making the fabrication of robust photonic insulators feasible. It has been shown that a small number of layers is enough to produce a sufficient attenuation. As a main conclusion we point-out that in the state of the art of thin-film technologies the fabrication of photonic insulator exhibiting total omni-directional external reflectivity within certain frequency region is realistic on the base of one-dimensional periodic structures. Recently, the predictions have been verified experimentally and good agreement with theory has been obtained. The experimental aspects of the problem will be the subject of forthcoming paper~\cite{Exp}.

Stimulating discussions with J.~Eggers and S.~Gaponenko are acknowledged. The authors thank J.~Eggers, S.~Rowson, J.~Broeng, V.~Kiselev and W.~Temnov for careful reading of the manuscript.

\begin{references}
\bibitem[\dagger]{DChigrin}
E-mail: chigrin@theo-phys.uni-essen.de

\bibitem{Basic}
E. Yablonovitch, Phys. Rev. Lett. {\bf 58},  2059 (1987); S. John, {\em 
ibid}. 2486 (1987).

\bibitem{PBGs}
See for example: E. Yablonovitch, J. Phys.: Condens. Matter {\bf 5},  2443 (1993); J.~B. Pendry, {\em ibid}. {\bf 8},  1085  (1996); J.~D. Joannopoulos, P.~R. Villeneuve, and S. Fan, Nature {\bf 386}, 143 (1997); {\em Photonic Band Gap Materials}, edited by C. Soukoulis (Kluwer Academic, Dordrecht, 1996); {\em Microcavities and Photonic Band Gaps: Physics and Applications}, edited by J.~G. Rarity and C. Weisbuch (Kluwer Academic, Dordrecht, 1996).

\bibitem{Joannopoulos}
J.~D. Joannopoulos, R.~D. Meade and J.~N. Winn, {\em Photonic crystals: molding the flow of light} (Princeton University Press, Princeton, N.J., 1995).

\bibitem{Bogomolov}
V.~N. Bogomolov {\em et al}., Appl. Phys. {\bf A63}, 613 (1996); M.~C. Wanke {\em et al}., Science {\bf 275}, 1284 (1997); E. {\" O}zbay {\em et al}., Opt. Lett. {\bf 19}, 1155 (1994); A. Rosenberg {\em et al}., Opt. Lett. {\bf 21}, 830 (1996). 

\bibitem{Patent}
D.~N. Chigrin and A.~V. Lavrinenko, in {\em Technical Digest of the 1998 OSA Annual Meeting and Exhibit}, (Baltimor, Maryland, USA, 1998), p. 118.

\bibitem{Born}
M. Born and E. Wolf, {\em Principles of Optics} (Pergamon, New York, 1980). 

\bibitem{Yeh}
P. Yeh, {\em Optical Waves in Layered Media} (John Wiley and Sons, New York, 1988).

\bibitem{Zolla}
F. Zolla, D. Felbacq and B. Guizal, Opt. Comm. {\bf 148}, 6 (1998); D. Felbacq, B. Guizal and F. Zolla, Opt. Comm. {\bf 152}, 119 (1998); J. Math. Phys. {\bf 39}, 4604 (1998).

\bibitem{Exp}
D.~N. Chigrin, A.~V. Lavrinenko, D.~A. Yarotsky, and S.~V. Gaponenko (to be published).

\end{references}
\end{document}